\documentclass[11pt]{article}
\usepackage{moriond,epsfig}
\def \bea{\begin{eqnarray}}
\def \beq{\begin{equation}}

\def \eea{\end{eqnarray}}
\def \eeq{\end{equation}}

\begin{document}
\rightline{Presented at XXXVIII Rencontres De Moriond:}
\rightline{Electroweak Interactions and Unified Theories}
\rightline{Les Arcs, France, March 15--22, 2003}
\vspace*{2cm}
\title{TOWARDS A MEASUREMENT OF $\phi_{3}$}

\author{S. K. SWAIN}

\address{Department of Physics and Astronomy,\\
 University of Hawaii at Manoa, \\
Honolulu, HI, USA}

\maketitle\abstracts{Results on the decays $B^{-} \rightarrow D_{CP}K^{-}$,
$\bar{B^{0}} \rightarrow D^{(*)0}\bar{K}^{(*)0}$, $B^{0} \rightarrow
D^{*\mp}\pi^{\pm}$ and their charge conjugates using data collected at the
$\Upsilon(4S)$ resonance with the Belle detector at the KEKB
asymmetric $e^{+}e^{-}$ storage ring are reported. The implications for the
determination of the weak phase $\phi_{3}$ are discussed.}

\section{$B^{-} \rightarrow D_{CP}K^{-}$}
The extraction of $\phi_{3}$~\cite{gamma}, an angle of the Kobayashi-Maskawa triangle~\cite{KM}, is a challenging measurement even with modern high luminosity $B$ factories. Recent theoretical work on $B$ meson dynamics has demonstrated the direct accessibility of $\phi_{3}$ using the process $B^{-} \rightarrow DK^{-}$~\cite{ADS,Gronau}. If the $D^{0}$ is reconstructed as a CP eigenstate, the $b \rightarrow c$ and $b \rightarrow u$ processes interfere. This interference leads to direct CP violation as well as a characteristic pattern of branching fractions. However, the branching fractions for $D$ meson decay modes to CP eigenstates are only of order 1~\%. Since CP violation through interference is expected to be small, a large number of $B$ decays is needed to extract $\phi_3$.  Assuming the absence of $D^0 - \bar{D^0}$ mixing, the observables sensitive to CP violation that are used to extract the angle $\phi_3$~\cite{BABAR} are,
\begin{eqnarray*}
{\cal{A}}_{1,2} \equiv \frac{{\cal B}(B^- \rightarrow D_{1,2}K^-) -
{\cal B}(B^+ \rightarrow D_{1,2}K^+) }{{\cal B}(B^- \rightarrow D_{1,2}K^-) + {\cal B}(B^+ \rightarrow D_{1,2}K^+) }&\\
= \frac{2 r \sin \delta ' \sin \phi_3}{1 + r^2 + 2 r \cos \delta ' \cos \phi_3}~~~~~~~~~~~~~~~~~~~~~~&\\
{\cal{R}}_{1,2} \equiv \frac{R^{D_{1,2}}}{R^{D^{0}}}  = 1 + r^2 + 2 r \cos \delta ' \cos \phi_3~~~~~~~~~~& \\
\delta ' = \left\{
             \begin{array}{ll}
              \delta & \mbox {{\rm  for }$D_1$}\\
              \delta + \pi&  \mbox{{\rm for }$D_2$}\\
             \end{array}
             \right.,~~~~~~~~~~~~~~~~~&
\end{eqnarray*}
where the ratios $R^{D_{1,2}}$ and $R^{D^{0}}$ are defined as
$$R^{D_{1,2}}=\frac{{\cal B}(B^- \rightarrow D_{1,2}K^-)+{\cal B}(B^+
\rightarrow D_{1,2}K^+)}{{\cal B}(B^- \rightarrow D_{1,2}\pi^-) +
{\cal B}(B^+ \rightarrow D_{1,2}\pi^+)},$$$$R^{D^{0}}=\frac{{\cal
B}(B^- \rightarrow D^{0} K^-)+{\cal B}(B^+ \rightarrow
\bar{D^{0}} K^+)}{{\cal B}(B^- \rightarrow D^{0}\pi^-) + {\cal B}(B^+
\rightarrow \bar{D^{0}} \pi^+)},$$$D_1$ and $D_2$ are CP-even and
CP-odd eigenstates of the neutral $D$ meson, $r$ denotes a ratio of amplitudes, $r \equiv |A(B^- \to \bar{D^0}K^-)/A(B^- \to D^0 K^-)|$, and $\delta$ is
their strong phase difference. Note that the asymmetries ${\cal{A}}_{1}$ and
${\cal{A}}_{2}$ have opposite signs. 
\begin{figure}[t]
\begin{center}
 \begin{tabular}{ll}
   \epsfig{file=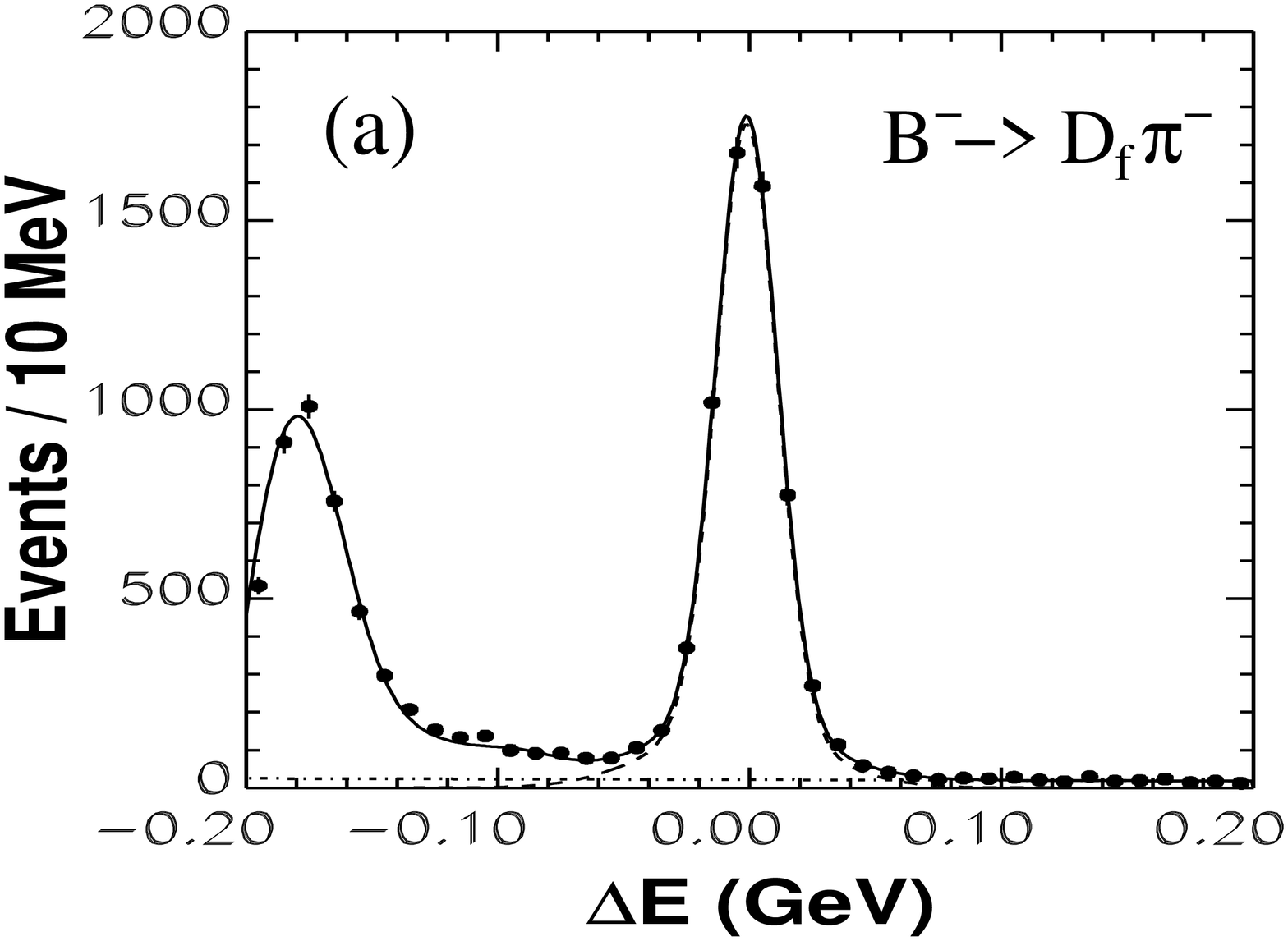,height=2.5cm,width=5.0cm} &
   \epsfig{file=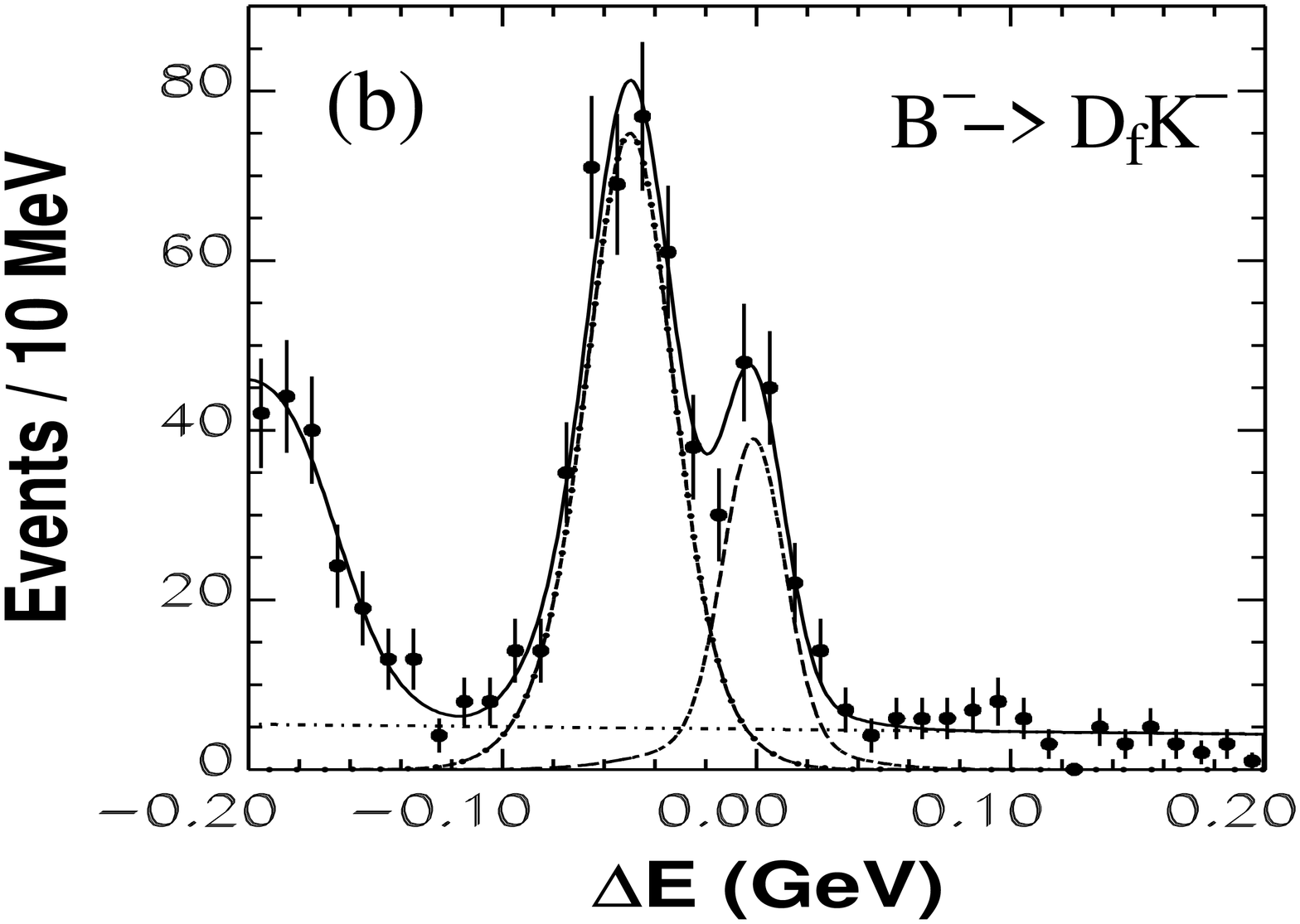,height=2.5cm,width=5.0cm}\\ 
   \epsfig{file=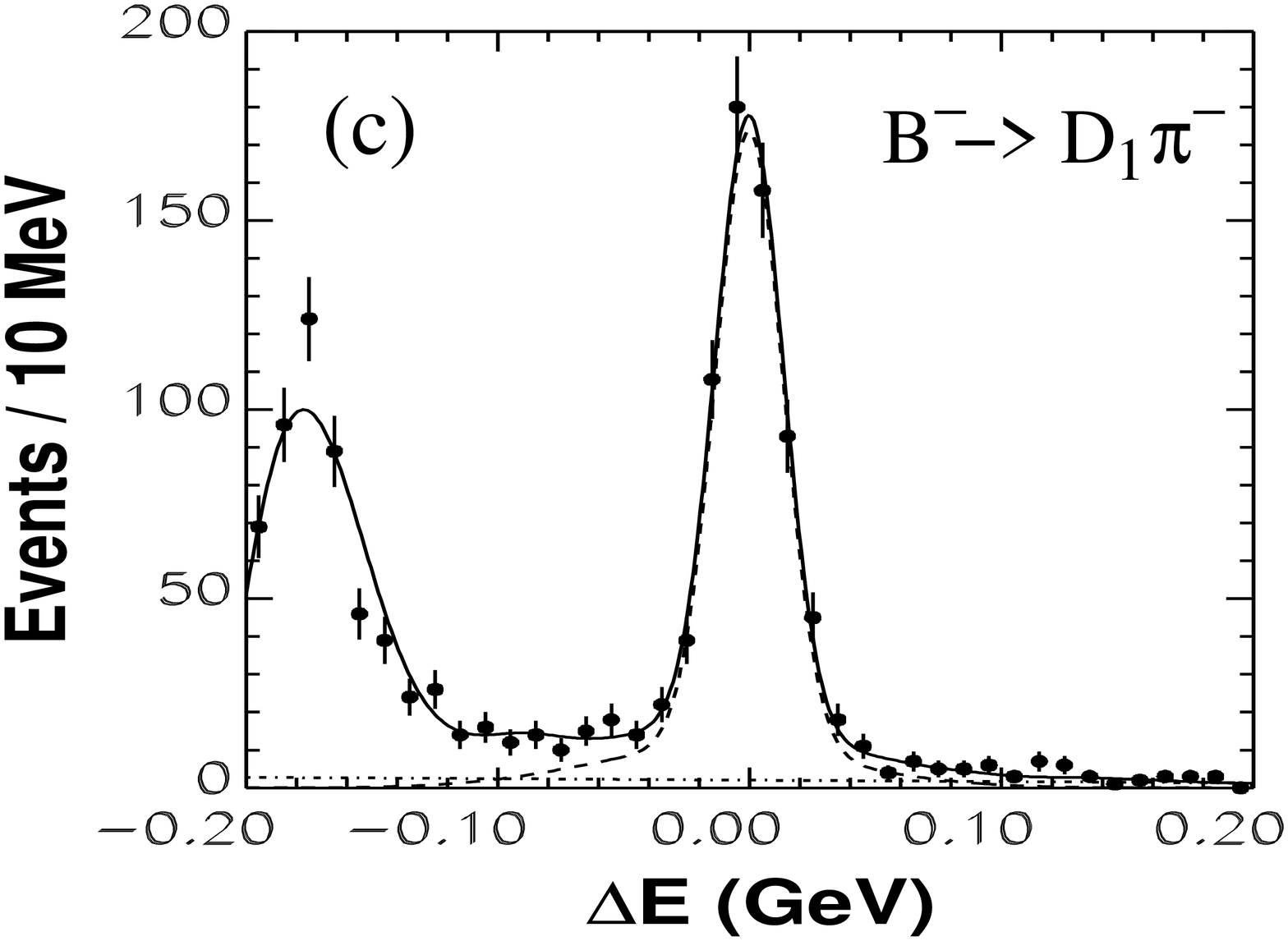,height=2.5cm,width=5.0cm} &
   \epsfig{file=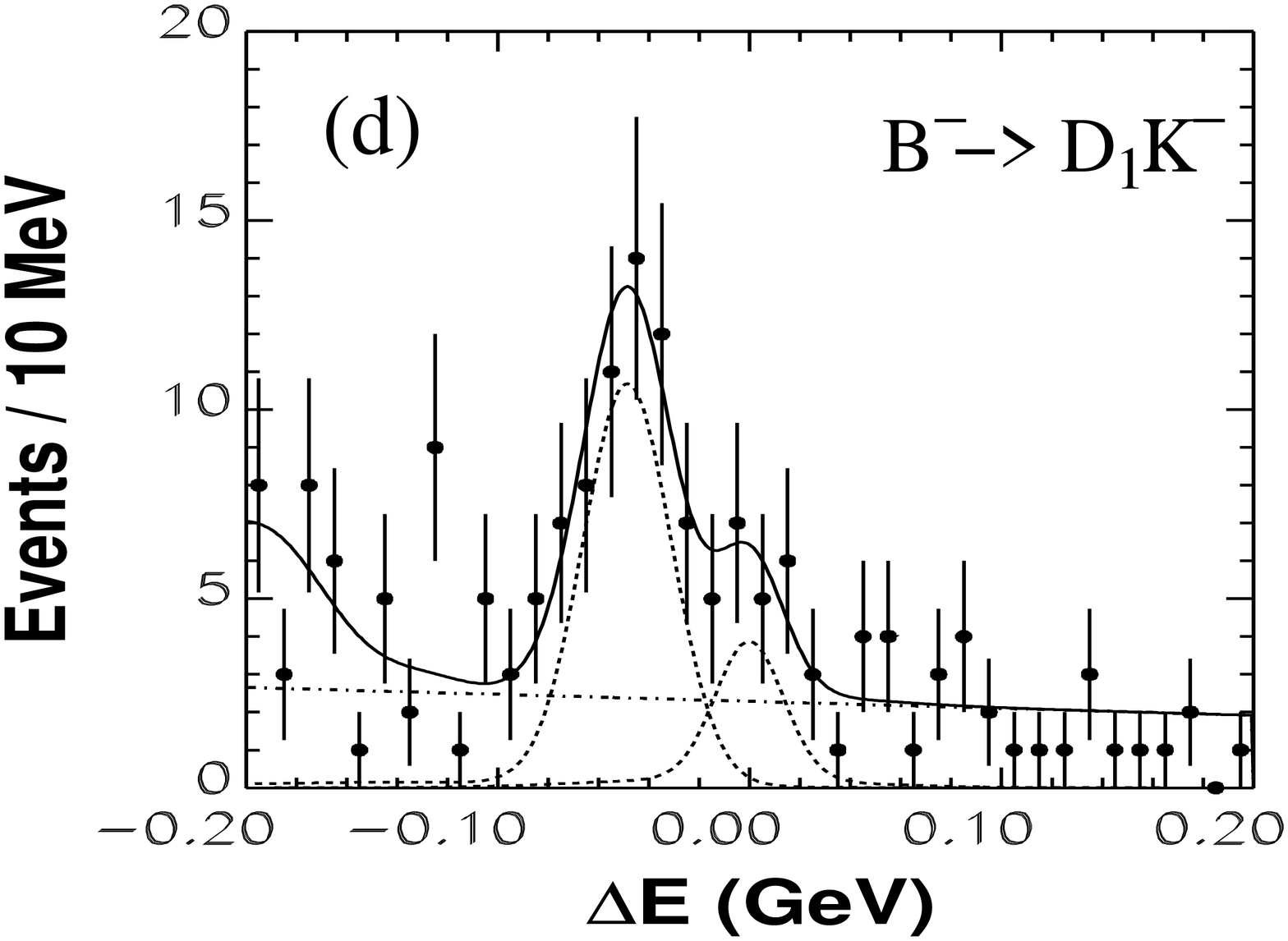,height=2.5cm,width=5.cm} \\
   \epsfig{file=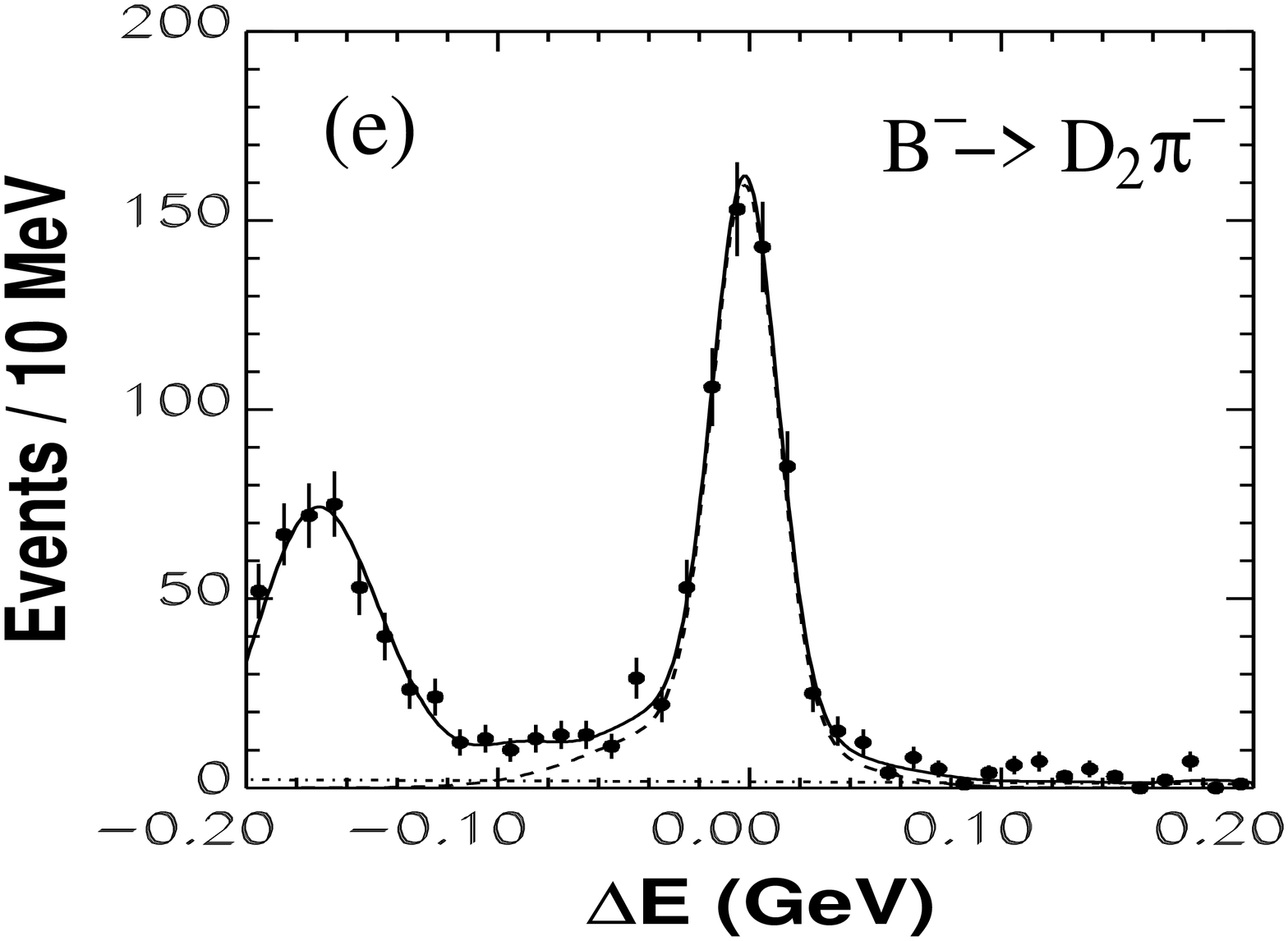,height=2.5cm,width=5.0cm} &
   \epsfig{file=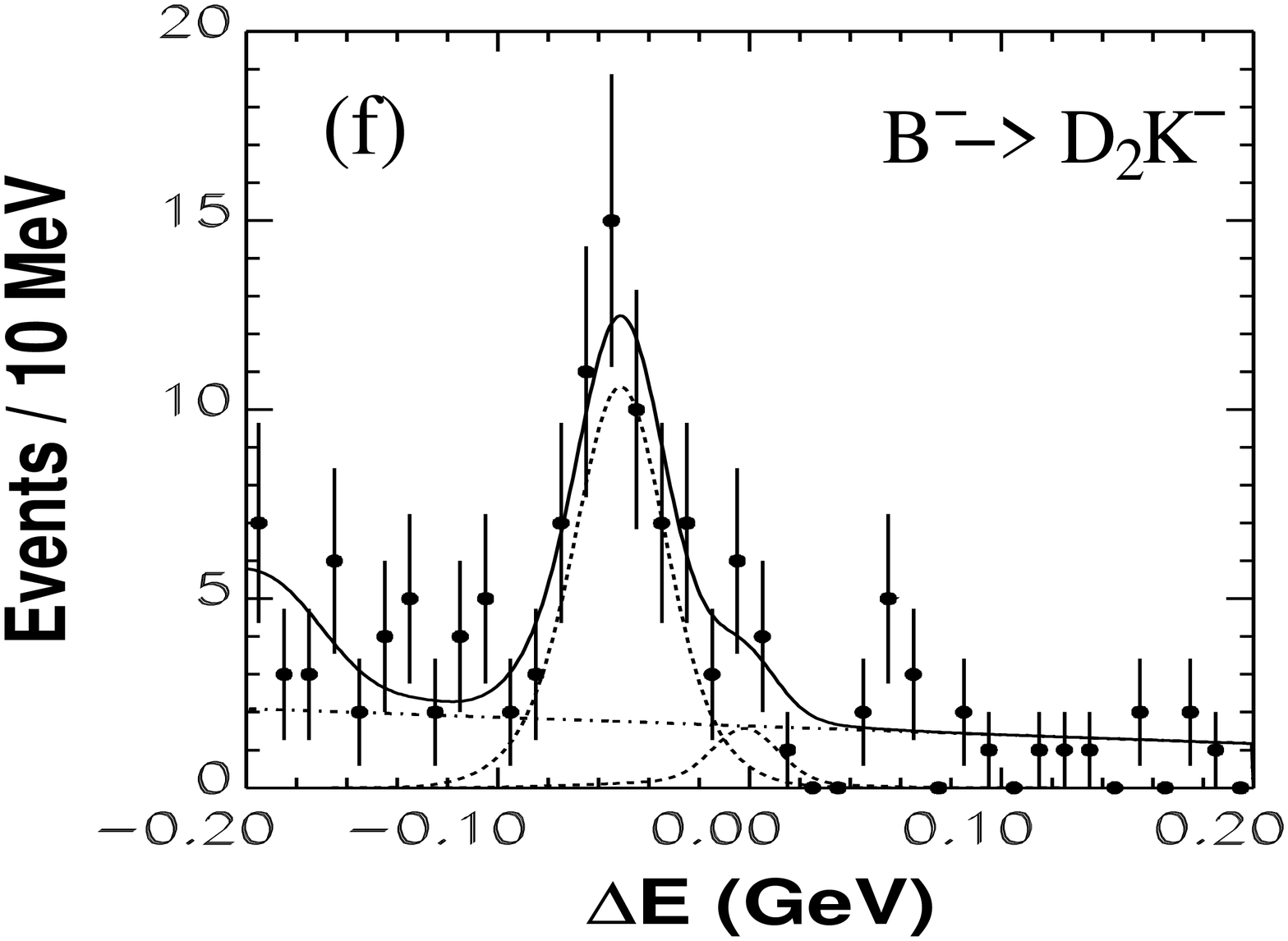,height=2.5cm,width=5.cm} \\
 \end{tabular}
\caption{$\Delta E$ distributions for (a)
$B^{-} \rightarrow D_{f}\pi^{-}$, (b)$B^{-} \rightarrow
D_{f}K^{-}$, (c)$B^{-} \rightarrow D_{1}\pi^{-}$,
(d)$B^{-} \rightarrow D_{1}K^{-}$, (e)$B^{-}
\rightarrow D_{2}\pi^{-}$ and (f)$B^{-} \rightarrow
D_{2}K^{-}$. Points with error bars are the data and the solid lines
show the fit results. \label{fig:dcpk}}
\end{center}
\end{figure}
\begin{table}
\begin{center}
\caption{Signal yields, feed-acrosses and ratios of branching fractions.
The errors on $R^{D}$ are statistical and systematic, respectively.\label{tab:tab1}}
\begin{tabular}{lcccc} \\ \hline
Mode & $B^- \rightarrow D \pi^-$ & $B^- \rightarrow D K^- $ & $B
 \rightarrow D \pi^-$& $R^{D} =\frac{{\cal B}(B^- \rightarrow D^0K^-)}{{\cal B}(B^- \rightarrow D^0\pi^-)}$  \\
 &  events               &  events              & feed-across  &  \\ \hline
$ B^- \rightarrow D_{f}h^{-}$ & 6052 $\pm$ 88 & 347.5 $\pm$ 21 & 134.4  $\pm$ 14.7& 0.077 $\pm$ 0.005 $\pm$ 0.006  \\
$ B^- \rightarrow D_{1}h^{-}$ & 683.4 $\pm$ 32.8 & 47.3 $\pm$ 8.9 & 15.6 $\pm$ 6.4 & 0.093 $\pm$ 0.018 $\pm$ 0.008  \\
$ B^- \rightarrow D_{2}h^{-}$ & 648.3 $\pm$ 31.0 & 52.4 $\pm$ 9.0 &
 6.3 $\pm$ 5.0 & 0.108 $\pm$ 0.019 $\pm$ 0.007  \\ \hline
\end{tabular}
\end{center}
\end{table}
\begin{table}
\begin{center}
\caption{Yields, partial-rate charge asymmetries and $90~\%$ C.L
intervals for asymmetries.\label{tab:tab2}}
\begin{tabular}{lccccc} \\ \hline
Mode & $N(B^{+})$ & $N(B^{-})$ & $\cal{A_{CP}}$ & $90~\%$ C.L \\ \hline
$ B^\pm \rightarrow D_{f}K^{\pm}$ & 165.4 $\pm$ 14.5 & 179.6 $\pm$ 15 & 0.04 $\pm$ 0.06$\pm$0.03 & $-$0.07$<{\cal A}_f<$0.15 \\
$ B^\pm \rightarrow D_{1}K^{\pm}$ & 22.1 $\pm$ 6.1 & 25.0 $\pm$ 6.5 & 0.06 $\pm$ 0.19 $\pm$0.04&  $-$0.26$<{\cal A}_1<$0.38 \\
$ B^\pm \rightarrow D_{2}K^{\pm}$ & 29.9 $\pm$ 6.5 & 20.5 $\pm$ 5.6 &
$-$0.19 $\pm$ 0.17$\pm$0.05 &   $-$0.47$<{\cal A}_2<$0.11  \\ \hline
\end{tabular}
\end{center}
\end{table}
 We reconstruct $D^{0}$ mesons in the following decay channels. For
the flavor specific mode (denoted by $D_{f}$), we use $D^{0}
\rightarrow K^{-}\pi^{+}$~\cite{CC}. For CP =+1 modes, we use $D_{1}
\rightarrow K^{-}K^{+}$ and $\pi^{-}\pi^{+}$ while for CP =$-$1 modes,
we use $D_{2} \rightarrow K_{S}^{0}\pi^{0}$, $K_{S}^{0}\phi$,
$K_{S}^{0}\omega$, $K_{S}^{0}\eta$ and $K_{S}^{0}\eta'$. We combine the $D^{0}$ and $\pi^{-}$/$K^{-}$ candidates (denoted by
$h$) to form $B$ candidates. The signal is
identified by two kinematic variables calculated in the center-of-mass
(c.m.) frame. The first is the beam-energy constrained mass,
$M_{\rm{bc}} = \sqrt{E_{\rm{beam}}^2 - |\vec{p}_D + \vec{p}_{h}|^2}$,
where $\vec{p}_{D}$ and $\vec{p}_{h}$ are the momenta of $D^{0}$ and
$K^-/\pi^{-}$ candidates and $E_{\rm{beam}}$ is the beam energy in the
c.m. frame. The second is the energy difference, $\Delta E = E_D +
E_{h} - E_{\rm{beam}}$, where $E_D$ is the energy of the $D^{0}$
candidate, $E_{h}$ is the energy of the $K^-/\pi^{-}$ candidate
calculated from the measured momentum and assuming the pion
mass,$~E_{h}=\sqrt{|\vec{p}_{h}|^2+ m_{\pi}^2}$. With this
definition, real $B^{-} \rightarrow D^{0}\pi^{-}$ events peak at
$\Delta E=0$ even when they are misidentified as $B^{-} \rightarrow
D^{0}K^{-}$, while $B^{-} \rightarrow D^{0}K^{-}$ events peak around
$\Delta E=-49$ MeV~\cite{dcpk}. The signal yields are extracted from a fit to the $\Delta E$
distribution in the region $5.27\,\mathrm{~GeV}/c^2 < M_{\rm{bc}} <
5.29\,\mathrm{~GeV}/c^2$. The fit results, using $78~\rm fb^{-1}$ data, are shown in Fig.\
\ref{fig:dcpk}. The signal yields and CP asymmetries are shown in
Table \ref{tab:tab1} and Table \ref{tab:tab2}.
\begin{figure}[t]
\begin{center}
 \begin{tabular}{ll}
   \epsfig{file=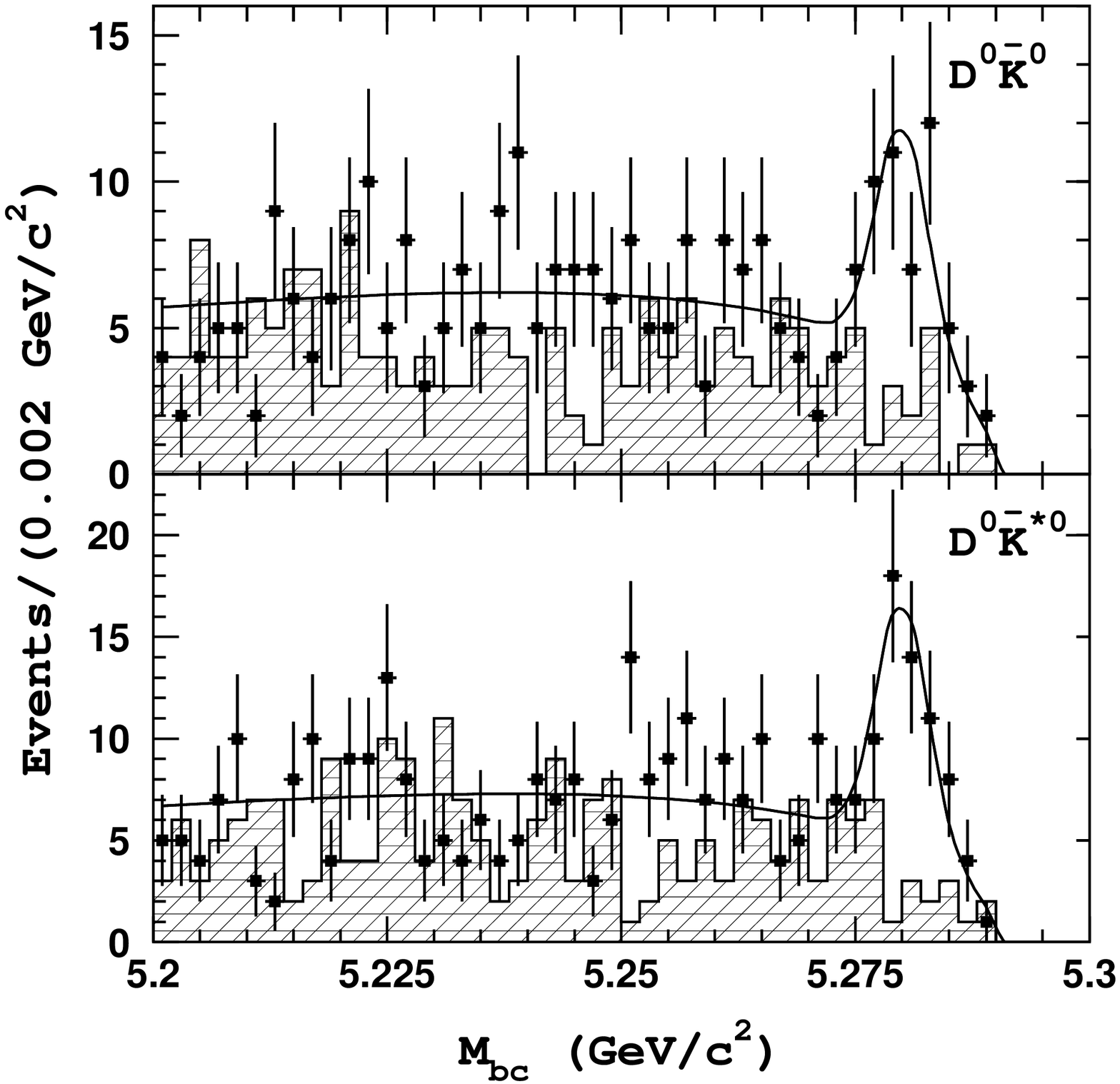,height=6cm,width=7.0cm} &
   \epsfig{file=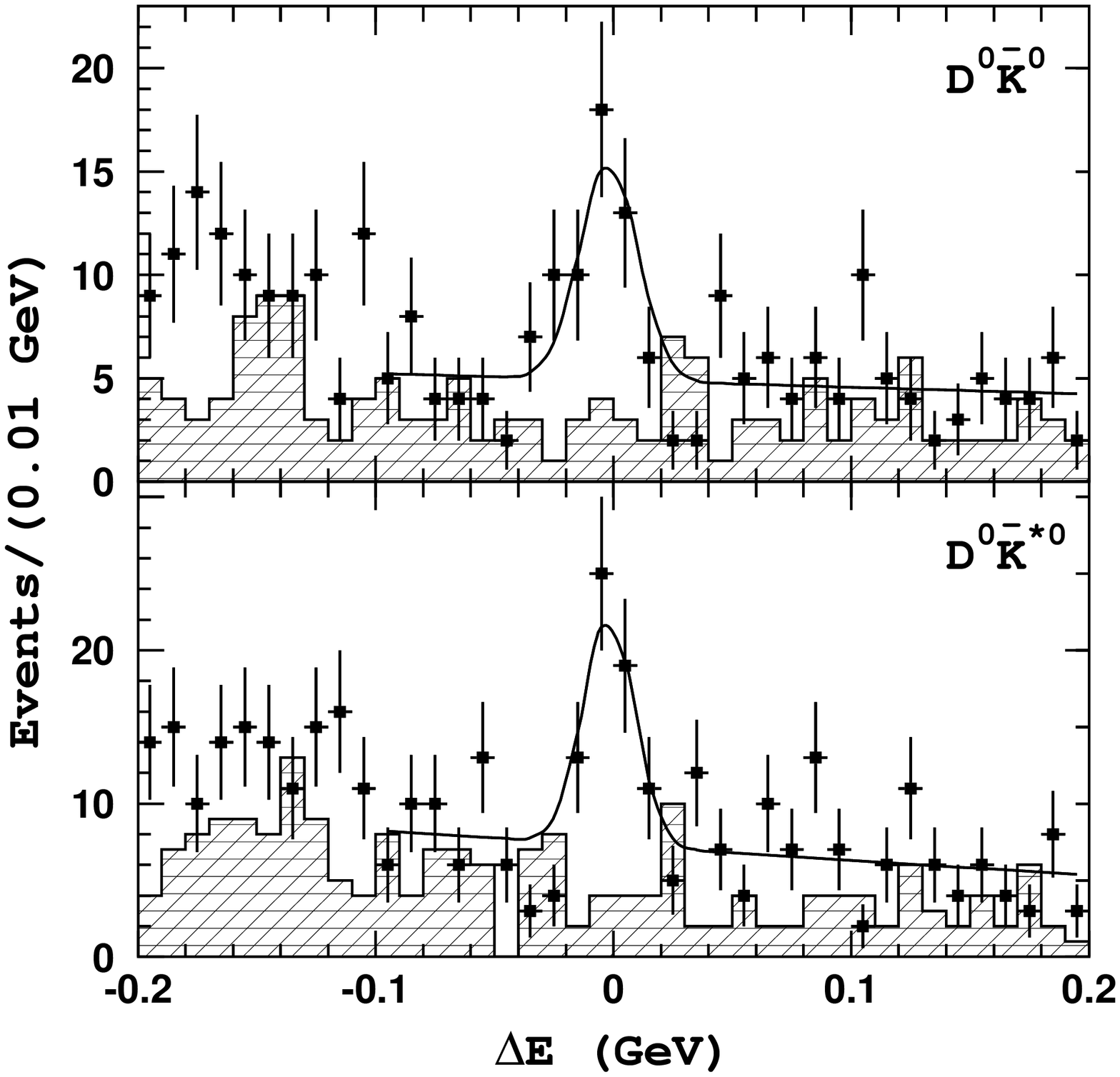,height=6cm,width=7.0cm}\\ 
 \end{tabular}
\caption{$\Delta E$(left) and $M_{bc}$(right) distributions for the
$\bar{B^{0}} \rightarrow D^{0}\bar{K}^{(*)0}$ candidates. Points with
errors represent the experimental data, hatched histograms show the
$D^{0}$ mass sidebands and curves are the results of the fits.\label{fig:dstkst}}
\end{center}
\end{figure}
\begin{table}[t]
\begin{center}
\caption{Fit results, branching fractions or upper limits at $90~\%$
C.L and statistical significances for $\bar{B^{0}} \rightarrow
\bar{D}^{*0}\bar{K}^{(*)0}$.\label{tab:tab3}}
\begin{tabular}{lccccc} \\ \hline
Mode & $\Delta E$ yield& $M_{bc}$ yield & ${\cal{B}}(10^{-5})$ &significance \\ \hline
$\bar{B^{0}} \rightarrow D^{0}\bar{K}^{0}$ &
$31.5^{+8.2}_{-7.6}$ & $27.0^{+7.6}_{-6.9}$ & $5.0^{+1.3}_{-1.2} \pm 0.6$  & $5.1\sigma$ \\
$\bar{B^{0}} \rightarrow D^{0}\bar{K}^{*0}$ & $41.2^{+9.0}_{-8.5}$ & $41.0^{+8.7}_{-8.1}$ & $4.8^{+1.1}_{-1.0} \pm 0.5$  & $5.6\sigma$ \\
$\bar{B^{0}} \rightarrow D^{*0}\bar{K}^{0}$ & $4.2^{+3.7}_{-3.0}$ &
$2.7^{+3.0}_{-2.4}$ & $<6.6$  & $1.4\sigma$ \\
$\bar{B^{0}} \rightarrow D^{*0}\bar{K}^{*0}$ & $6.1^{+5.2}_{-4.5}$ &
$8.6^{+4.2}_{-3.6}$ & $<6.9$  & $1.4\sigma$ \\
$\bar{B^{0}} \rightarrow \bar{D}^{0}\bar{K}^{*0}$ &
$1.4^{+8.2}_{-7.6}$ & $9.2^{+7.7}_{-7.2}$ & $<1.8$  & $-$ \\
$\bar{B^{0}} \rightarrow \bar{D}^{*0}\bar{K}^{*0}$ & $1.2^{+4.1}_{-3.6}$ & $0.0^{+3.9}_{-3.2}$ & $<4.0$ & $-$ \\ \hline
\end{tabular}
\end{center}
\end{table}
\section{$\bar{B^{0}} \rightarrow D^{(*)0}\bar{K}^{0}$ and  $\bar{B^{0}} \rightarrow D^{(*)0}\bar{K}^{*0}$}
The two-body decays of the above type, which occur via tree-level
diagrams, can be used to test the factorization hypothesis. Precise measurements of the
decay rates allow one to construct the isospin relation between the 
transition amplitudes  and determine  the relevant strong and weak
phase. The modes $\bar{B^{0}} \rightarrow
D^{0}\bar{K}^{*0}$, $\bar{B^{0}} \rightarrow \bar{D}^{0}\bar{K}^{*0}$
and $\bar{B^{0}} \rightarrow D_{CP}\bar{K}^{*0}$ decays also allow a measurement of the angle
$\phi_{3}$. We reconstruct $D^{0}$ mesons in the decay channels:
$K^{-}\pi^{+}$, $K^{-}\pi^{+}\pi^{0}$ and $K^{-}\pi^{+}\pi^{-}\pi^{+}$, using
a requirement that the invariant mass be within $20\,\mathrm{MeV}/c^2$,
$15\,\mathrm{MeV}/c^2$ and $25\,\mathrm{MeV}/c^2$ of the nominal
$D^{0}$ mass, respectively. In each channel we further define a
$D^{0}$ mass sideband region, with width twice that of signal
region. For the $\pi^{0}$ from the $D^{0} \rightarrow
K^{-}\pi^{+}\pi^{0}$ decay, we require that its momentum in the CM
frame be greater than $0.4\,\mathrm{GeV}/c$ in order to reduce
combinatorial background. $D^{*0}$ mesons are reconstructed in the $D^{*0} \rightarrow
D^{0}\pi^{0}$ decay mode. The mass difference between $D^{*0}$ and
$D^{0}$ candidates is required to be within $4\,\mathrm{MeV}/c^2$ of
the expected value. $\bar{K}^{*0}$ candidates are reconstructed from
$K^{-}\pi^{+}$ pairs with an invariant mass within
$50\,\mathrm{MeV}/c^2$ of the nominal $\bar{K}^{*0}$ mass. We then
combine $D^{(*)0}$ candidates with $K^{0}_{S}$ or $\bar{K}^{*0}$ to
form B mesons. For the final result using $78~\rm fb^{-1}$ data, a simultaneous fit to the
$\Delta E$ distributions for the three $D^{0}$ decay channels taking
into account the corresponding detection efficiencies~\cite{dstk}. The fit result
is shown in Fig.\ \ref{fig:dstkst}. The signal yields from the fitting
and the branching fractions are shown in Table \ref{tab:tab3}. 
\begin{figure}[t]
\begin{center}
 \begin{tabular}{ll}
   \epsfig{file=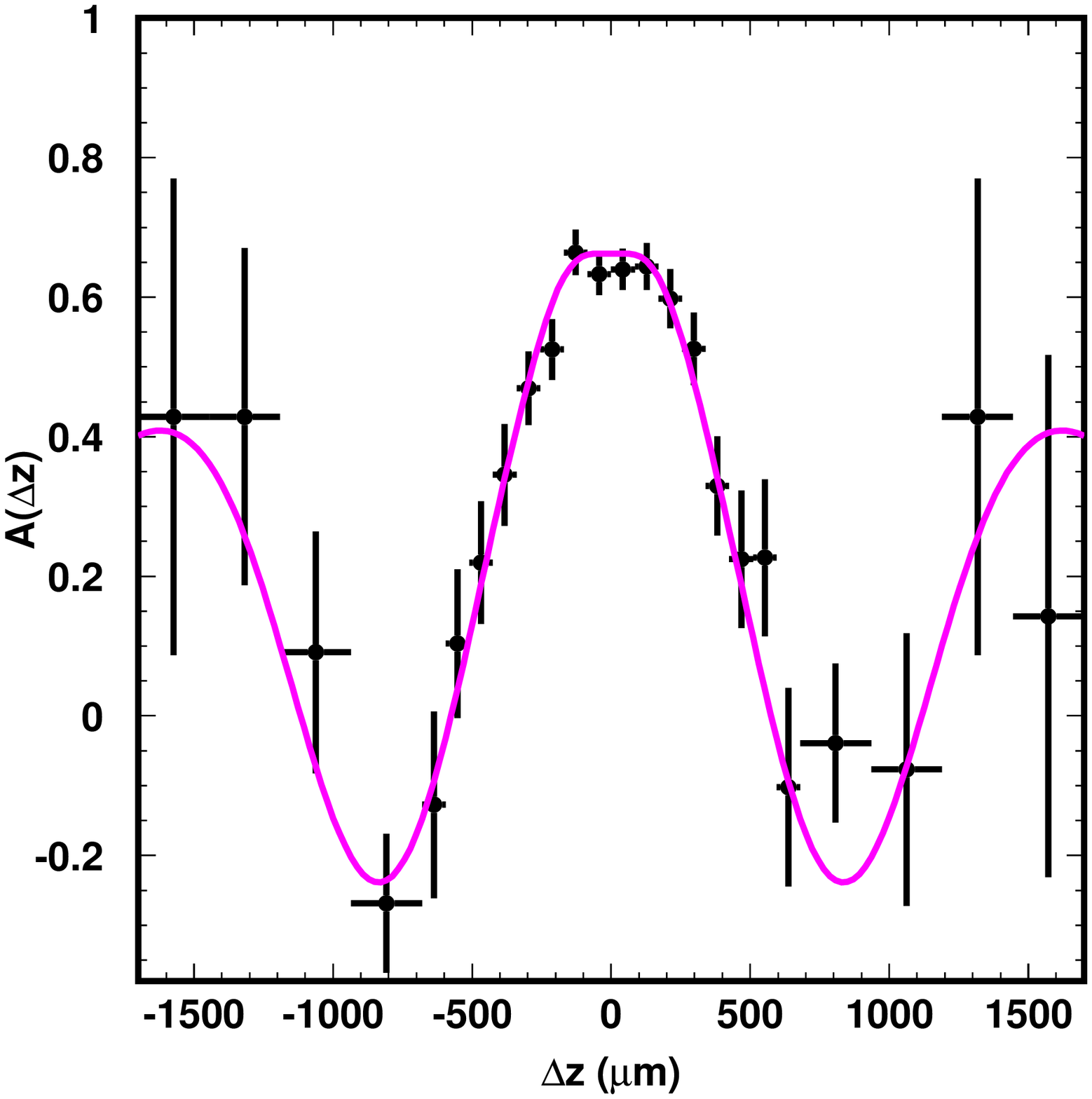,height=4.3cm,width=6.0cm} &
   \epsfig{file=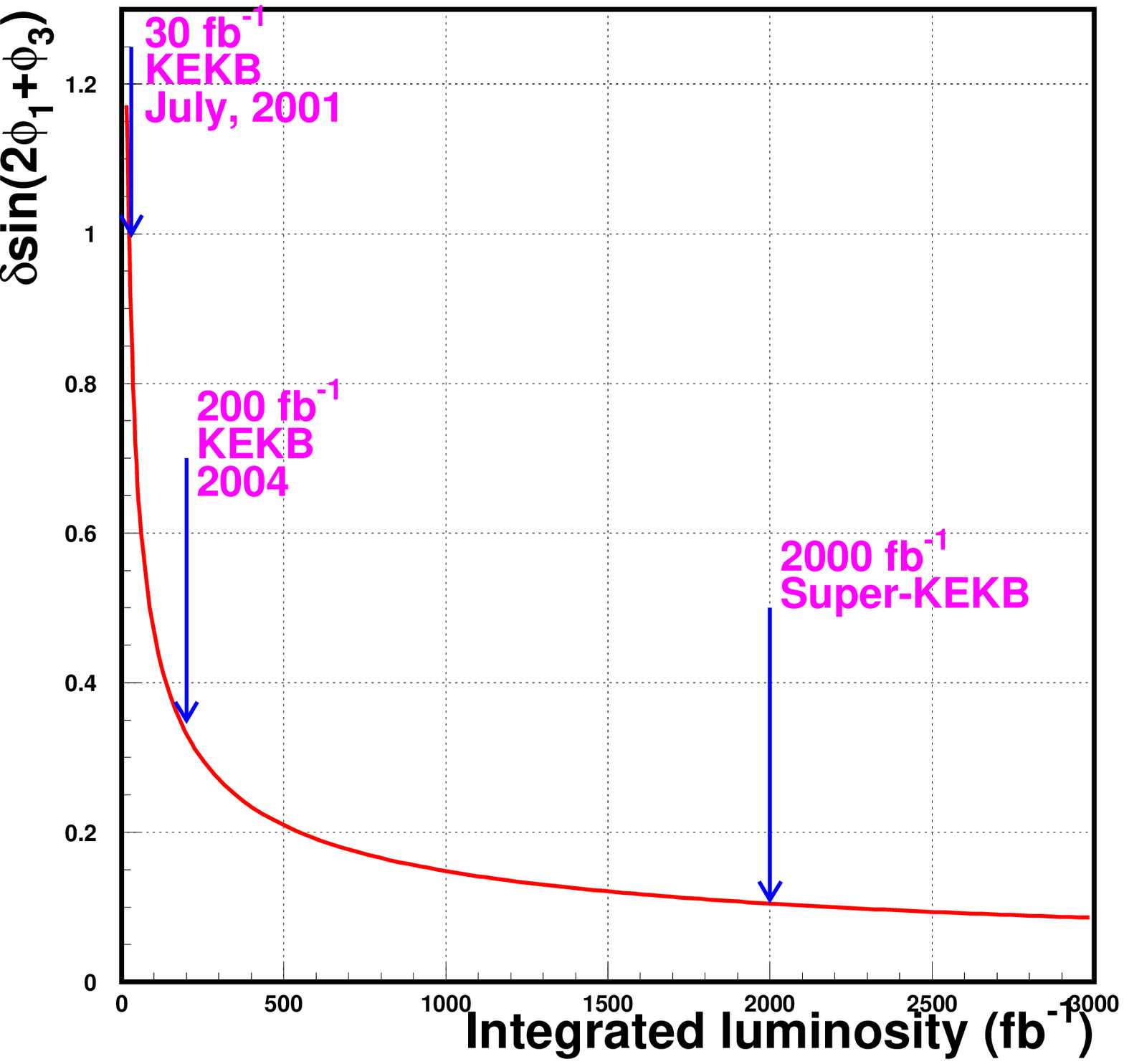,height=4.5cm,width=6.0cm}\\ 
 \end{tabular}
\caption{(Left)Distribution of the asymmetry, $A(\Delta z)$, as a function
of $\Delta z$ for the data with the fit curve overlaid. (Right) Error
on $\rm sin(2\phi_1+\phi_3)$, as a function of integrated luminosity.\label{fig:mixing} }
\end{center}
\end{figure}
\section{$B^{0}-\bar{B}^{0}$ mixing with $B^{0}(\bar{B}^{0})
\rightarrow D^{*\mp}\pi^{\pm}$ partial reconstruction.}
Since both Cabibbo-favoured ($B^0 \rightarrow D^{*-}\pi^+$) and
Cabibbo-suppressed ($\bar{B}^0 \to D^{*-}\pi^+$) decays contribute to the
$D^{*-}\pi^+$ final state, a time-dependent analysis can be used to
measure $\rm sin(2\phi_1+\phi_3)$.  Since the ratio of amplitudes is
expected to be small ($\sim0.02$), the CP asymmetry will be hard to
observe, but may be possible since the $B^0 \rightarrow D^{*-} \pi^+$ decay rate
is fairly large.  A first step towards this measurement is the extraction of
the mixing parameter $\Delta m_d$ from $B^0 \to D^{*-} \pi^+$.

We use events with a partially reconstructed $B^{0}(\bar{B}^{0})
\rightarrow D^{*\mp}\pi^{\pm}$ candidates and where the flavor of the
accompanying $B$ meson is identified by the charge of the lepton from
a $B^{0}(\bar{B}^{0}) \rightarrow X^{\mp}l^{\pm}\nu$ decay. The
proper-time difference between the two $B$ mesons is determined from
the distance between the two decay vertices ($\Delta Z$). From a simultaneous fit
to the proper-time distributions for the same flavor(SF) and opposite flavor(OF)
event samples, we measure the mass difference between the two mass
eigenstates of the neutral $B$ meson to be $\Delta m_{d} = (0.509 \pm
0.017(stat) \pm 0.020(sys)) ps^{-1}$. The result is obtained using
$29.1~\rm fb^{-1}$ data collected with Belle detector at KEKB. This is the
first direct measurement of $\Delta m_{d}$ using the technique of
partial reconstruction. Fig.\ \ref{fig:mixing}(left) shows the mixing asymmetry $A(\Delta Z)$ as a function of $\Delta Z$ where \\
\beq
A(\Delta Z) \equiv  \frac{N^{OF}(\Delta Z) - N^{SF}(\Delta
Z)}{N^{OF}(\Delta Z) + N^{SF}(\Delta Z)}
\eeq
 where $N(\Delta Z)$ is the yield of the signal candidates as a
function $\Delta Z$~\cite{dstpi}. This method can be extended to measure the weak
angle $2\phi_{1} + \phi_{3}$. The expected statistical error on $\rm sin(2\phi_1+\phi_3)$ is
estimated from a large Monte Carlo sample that does not include the
effects of backgrounds and mistagging is shown in Fig.\ \ref{fig:mixing}(right).  The
expected sensitivity is around 0.35 at $200\ {\rm fb}^{-1}$.

\section*{Acknowledgments}
We wish to thank the KEKB accelerator group for the excellent
operation of the KEKB accelerator.


\begin{thebibliography}{99}
\bibitem{gamma} Another naming convention, 
$\gamma$ ($= \phi_3$), is
also used in the literature.
\bibitem{CC}
Hereafter, the inclusion of the charge conjugate mode
decay is implied unless otherwise stated.
\end{thebibliography}
\end{document}